\documentclass[proof]{rmaa}

\usepackage{paralist}

\usepackage{psfrag,color}

\usepackage[latin1]{inputenc}

\title{Metallicity gradients in M31, M\,33,  NGC 300 and the Milky Way using abundances of different elements} 
\author{
  Miriam Pe\~na\altaffilmark{1} 
  and Sheila N. Flores-Dur\'an\altaffilmark{1}}
\altaffiltext{1}{Instituto de Astronom\'\i{}a,
  Universidad Nacional Aut\'onoma de M\'exico,
   Apdo. Postal 70264, Ciudad de M\'exico, 04510, M\'exico.}

\shortauthor{Pe\~na, \& Flores-Dur\'an}
\shorttitle{Metallicity gradients}

\fulladdresses{
\item S. N. Flores-Dur\'an: Instituto de Astronom{\'\i}a,
  Universidad Nacional Aut\'onoma de M\'exico,
   Apdo. Postal 70264, Ciudad de M\'exico, 04510, M\'exico).
\item M. Pe\~na: Instituto de Astronom{\'\i}a,
  Universidad Nacional Aut\'onoma de M\'exico,
   Apdo. Postal 70264, Ciudad de M\'exico, 04510, M\'exico
  (miriam@astro.unam.mx).}

\listofauthors{M. Pe\~na \& S. N. Flores-Dur\'an}
\indexauthor{Pe\~na, M.}
\indexauthor{Flores-Dur\'an, S. N.}
\abstract{Metallicity gradients derived from planetary nebulae (PNe) using O, Ne, and Ar abundances are studied and compared to those from H\,{\sc ii} regions in the  galaxies M\,31, M\,33, NGC\,300 and  the Milky Way.  Galactocentric radii and chemical abundances were collected from the literature, carefully selecting a homogeneous sample for each galaxy. Metallicity gradients shown by PNe are flatter than those of H\,{\sc ii} regions in all cases. The extreme case is M\,31 where  PN abundances  are not related to galactocentric distances and the gradients are consistent with {\bf zero}. To analyze the evolution of gradients with time we build gradients for Peimbert Type I  and non-Type I PNe  finding that Type I PNe show steeper gradients than non-Type I PNe and more similar to the ones of H\,{\sc ii} regions indicating that the chemical gradients might steepen with time. Alternatively, the flat gradients for old PNe show that radial migration could have an important role in the evolution of galaxies.}

\resumen{Estudiamos los gradientes de metalicidad de O, Ne y Ar, derivados de nebulosas planetarias (PNe), en comparaci\'on con los  de regiones H\,{\sc ii} en las galaxias M\,31, M\,33, NGC\,300 y la V{\'\i}a L\'actea. Radios galactoc\'entricos y abundancias fueron recopilados de la literatura, seleccionando con cuidado una muestra 
homog\'enea de objetos en cada galaxia. Los gradientes de las PNe son m\'as planos que los de las regiones H\,{\sc ii} en todos los casos. El caso m\'as extremo es el de M\,31, donde las abundancias de las PNe no est\'an relacionadas con la distancia galactoc\'entrica y los gradientes son consistentes con cero. Calculamos gradientes para PNe del Tipo I y no-Tipo I de Peimbert, encontrando que los gradientes del Tipo I son m\'as empinados y m\'as similares a los de las regiones H\,{\sc ii}, lo que indicar{\'\i}a que los gradientes de metalicidad se empinan con el tiempo. Alternativamente los gradientes planos de las viejas PNe indican que la migraci\'on radial juega un importante papel en las galaxias. }

\addkeyword{galaxies: spiral }
\addkeyword{galaxies: ISM: abundances}
\addkeyword{planetary nebulae: abundances}
\addkeyword{Galaxies: individual: M 31, M 33, NGC\ 300 }
\addkeyword{The Galaxy}

\begin{document}
% Typeset article header
\maketitle
\section{Introduction}
\label{sec:intro}

Metallicity gradients in disk galaxies, provided by the analysis of the chemistry of H\,{\sc ii}  regions at different galactocentric distances, have been studied since long ago \citep[and many others]{aller:42, searle:71}, as such an analysis gives information on the chemical history of the host galaxy. The history of star formation and the processes of accretion and mass loss in a galaxy can be determined by using chemical evolution models that reproduce the present  chemical abundances of the interstellar medium (ISM)  using the abundances of H\,{\sc ii} regions as constraint. See e.g., \citet{carigi:11}.

It has been found that metallicity gradients obtained from H\,{\sc ii} regions and other indicators are always negative, that is, chemical abundances are lower at larger galactocentric distances. Many galaxies have been analyzed showing the same result.   Recently \citet{sanchez:14} analyzed the O abundance in a large number of H\,{\sc ii} regions in more than 300 galaxies observed by  the  CALIFA survey, reporting that in many galaxies with H\,{\sc ii} regions detected beyond two disk effective radii, the slope  presents a flattening at large distances and in some cases a drop or truncation of the O abundance occurs in the inner regions.

Since some time ago  the abundances determined for PNe are being also used to analyze  the chemical gradients in galaxies.  Among the oldest studies of PNe in the Milky Way we find that by \citet{dodorico:76} and \citet{kaler:80}. As PNe are objects with ages between 1 and 10 Gyr, they provide information of the past ISM abundances, helping then in the determination of the evolution of the chemical abundances in a galaxy, and providing additional constraints for the chemical evolution models of a galaxy \citep{hernandez:11}.

Oxygen is the most used element to determine the chemistry of H\,{\sc ii} regions because its  abundance in these photoionized nebulae is well determined by just adding the ionic abundances of O$^+$ and O$^{++}$, whose lines are observed in the visual range of wavelengths. The determination of ionic abundances requires the determination of physical conditions in the plasma, in particular the electron temperature which can be derived also in the visual range by detecting the faint auroral lines [O{\sc iii}] $\lambda$4363 and [N{\sc ii}] $\lambda$5755, in order to use the so called {\it direct-method} to determine abundances. 

Gradients based on oxygen abundances have been also calculated from PN data although for the case of highly ionized objects a correction due to the presence of O$^{+3}$ is required. However, it has been  shown that O in PNe may be enriched (or  depleted) due to stellar nucleosynthesis, particularly in low-metallicity environments \citep{pena:07,flores-duran:17} and also in Galactic PNe with carbon-rich dust \citep{delgado-inglada:15}.
 Therefore metallicity gradients derived from O abundances in PNe could be perturbed by stellar nucleosynthesis and thus this element  could be not adequate to analyze the chemical gradients and the evolution of galaxies. Apparently Ne  abundances are also modified by stellar nucleosynthesis in low-metallicity environments \citep{karakas:10,milingo:10,flores-duran:17}.  On the other hand, Ar and S abundances in PNe are not expected to be modified in such processes during the PN progenitor lifetime, although their abundance determinations have large uncertainties due to large uncertainties in the ionization correction factors when only one ion (Ar$^{++}$ or S$^+$) is observed \citep{delgado-inglada:14}. Especially S abundance presents several problems, like 'the Sulphur anomaly'. See e.g., \citet{henry:10}.

Considering the above, in this work we propose to use Ar/H, together with O/H and Ne/H abundance ratios,  to trace the PN metallicity gradients in the galaxies of the local universe M\,31, M\,33, NGC\,300, and the Milky Way (MW). Gradients of PNe will be discussed in comparison with those of H\,{\sc ii} regions.  In \S 2 we present the data used for the different galaxies. In \S 3 the abundance gradients are calculated for each galaxy. In \S 4 our results are presented and discussed, and our conclusions are indicated in \S 5.

\section{Data Acquisition}
\label{data}

Data used in this work consist of abundances and galactocentric distances  of PNe and H\,{\sc  ii} regions of the spiral galaxies M\,31, M\,33, NGC\,300, and the Milky Way. Most data were obtained from the literature trying to include only abundances calculated with the {\it direct-method} (electron temperature determined), in a homogeneous way by one group of authors. However in some cases data from different sources have been included in order to obtain a larger sample of objects  for statistical studies, covering as much galactic radius as possible. This introduced some degree of inhomogeneity  that has been minimized by adopting  only objects with electron temperature determined and by using recent ionization correction factors. We have selected samples where O, Ne, and Ar abundances are available.

% In the case of M\,31 we collected data of PNe  from different authors, therefore we used the extinction-corrected line intensities published by them to calculate the chemical abundances in a homogeneous way (see \S 2.1). 

We intend to build gradients based only on data from genuine PNe and H\,{\sc ii} regions, therefore for all the galaxies, we carefully selected as genuine PNe those objects where [O{\sc iii}] $\lambda$5007 intensity,  relative to H$\beta$ is larger than 3, because a lower value  could correspond to a compact H\,{\sc ii} region and not to a bona-fide PN. This criterium has been used for instance by \citet{ciardullo:02}  to select PNe to build the PNLF in external galaxies. On the other hand,  H\,{\sc ii} regions showing [O{\sc iii}] $\lambda$5007intensity  larger than 3 times H$\beta$ were eliminated from the sample because these objects could correspond to  nebulae around WR stars or SN remnants, whose abundances can be contaminated by the processes in the central stars and would not correspond to abundances of authentic pre-star nebulae.  

In Table \ref{tab:R25} we present several characteristics of the studied galaxies such as their Hubble type, mass, distance to the Milky Way and the optical radius $R_{25}$ (the 25 mag arcsec$^{-2}$ isophotal radius).

\begin{table}
\caption{Characteristics of galaxies}
	\label{tab:R25}
	\begin{tabular}{llrlrrl}
	\hline
	Name & Hubble & Mass & Dist& O/H$^{(a)}$ & $R_{25}$  & ref. for \\
 	&Type &$M_\odot$ & (kpc) & & (kpc) &  $R_{25}~^{(b)}$\\
	\hline
	M\,31 & SA(s)b&1.5E12 & 785 &8.8 & 20.6 & D14%\citet{draine:14}
 \\	
MW & SBbc& 8.0E11 &--- &8.8  & 11.5 &  S-M18%\citet{sanchezm:18} 
\\
	M\,33 & SA(s)cd & 5.0E10 &849 & 8.5 &9.0 & M07%\citet{magrini:07}
 \\
	NGC\,300 &SA(s)d & 3.5E10&1880  &8.6 & 5.3 & B09%\citet{bresolin:09} 
\\

%	M81 & 14.6 & Patterson et al. (2012) \\
	\hline
\multicolumn{7}{l}{$^{(a)}$ O/H is the abundance at R=0, 12+log O/H,  from H\,{\sc II} regions.}\\
\multicolumn{7}{l}{$^{(b)}$ D14: \citet{draine:14}, M07: \citet{magrini:07}  }\\
\multicolumn{7}{l}{B09: \citet{bresolin:09}, S-M18: \citet{sanchezm:18}}\\
	\end{tabular}
\end{table}

\subsection{M\,31}
\label{sec:m31} % used for referring to this section from elsewhere

The well known M\,31 is the most massive spiral galaxy in the Local Group. It is at a distance of 785 kpc  from the Milky Way \citep{mcconnachie:05}.

Chemical gradients calculated from the abundances of H\,{\sc ii} regions  have been studied by several authors, most of the times based on abundances derived by using {\it strong-line} methods because due to the high metallicity in this galaxy, the auroral lines indicative of electron temperature, such as [O{\sc iii}]$\lambda$4363 and  [N{\sc ii}]$\lambda$5755, are faint and  difficult to detect. The biggest sample analyzed in this way is the one by  \citet{sanders:12}, which included 192 H\,{\sc ii} regions. They reported an O abundance gradient of about $-0.0195\pm0.0055$ dex kpc$^{-1}$, but found that the slope depends on the choice of the {\it strong-line} method used.

 \citet{zurita:12} obtained abundances based on the {\it direct-method} for 31 H\,{\sc ii} regions concentrated at two galactocentric distances of 3.9 and 16.1 kpc. These authors discussed the O/H abundance gradient finding a robust negative slope of $\Delta$O/H / $\Delta$R = $-$0.023 dex kpc$^{-1}$ based on both, the {\it direct-method} and the {\it strong-line} method.  This value is similar to the values reported by other authors, based on {\it strong-line} methods (Zaritsky, Kennicutt, \& Huchra 1994; Sanders et al. 2012). 

Interestingly, \citet{zurita:12} found that O abundances, determined with the {\it direct-method} in their H\,{\sc ii} regions,  are lower,  by about  0.3 dex, than the values determined with the {\it strong-line} methods and the values derived for supergiant stars, at any galactocentric distance. The authors attribute  this discrepancy to a bias in their sample which would be not representative of the mean H\,{\sc ii} region population (because only high-temperature regions, and thus of low metallicity,  could be detected in their search), and to the probable depletion of O in dust grains.
 Abundance data for H\,{\sc ii} regions, used in this work,  are taken  from  \citet{zurita:12}.

Data for PNe were collected from the articles by \citet{kwitter:12}; \citet{sanders:12}; \citet{balik:13}; and \citet{fang:13, fang:15}.  The use of data from different authors can introduce undesirable inhomogeneities, therefore, in order to analyze chemical abundances on a more homogeneous  system, physical conditions and ionic abundances of PNe  were recalculated by us from the extinction-corrected line intensities published by the cited authors. The IRAF 2.16 five level nebular modelling package "stsdas.analysis.nebular", with the  tasks {\it temden} and {\it ionic}, were employed to determine physical conditions and ionic abundances. 
 For all the sample the [\ion{O}{iii}] temperature was derived and occasionally the [\ion{N}{ii}] temperature was available, but all the ionic abundances were calculated with the [\ion{O}{iii}] temperature (one-temperature zone model).
Total abundances were  derived from the ionic abundances  by using the ionization correction factors (ICFs) by \citet{delgado-inglada:14} to correct for the  not seen ions. The results, although not too different from the ones published by the cited authors, are now in a homogeneous system.

 The PN distribution in this galaxy extends up to $R/R_{25}\sim 5$ (R $\geq$  100 kpc) while data for the H\,{\sc ii} regions cover only up to $R/R_{25}\sim 1$ (R $\sim$ 20 kpc). 

The galactocentric distances used in the diagrams for the gradients  were collected from the same references used for the abundance ratios.

\subsection{M\,33}

M\,33 is the third spiral galaxy in the Local Group. Its Hubble type is very late, SA(s)cd, and its mass is much lower  than the masses of the Milky Way and M\,31, by factors of 16 and 30 respectively. It is located at 849 kpc from the Milky Way and appears almost face-on. The central metallicity in M\,33 as given by H\,{\sc ii} regions,  12+log O/H= 8.5,  is lower than in the bigger spirals, and it  is even lower than in NGC\,300 (see Table \ref{tab:lin-fit}).

Gradients from PN abundances are found in \citet{magrini:09} and \citet{bresolin:10} who found that the slopes are   equal to the ones of H\,{\sc ii} regions or flatter, respectively.

For this galaxy, PN data were collected from \citet{bresolin:10} and \citet{magrini:10}, and  H\,{\sc ii} region data are from \citet{magrini:10}. 
 The distributions of H\,{\sc ii} regions and PNe   extend up to $R/R_{25}\simeq 1$ (R$\sim$ 9 kpc).  As said in the Introduction, in the sample of PNe we excluded those objects showing [O{\sc iii}]$\lambda$5007/H$\beta$ intensity ratio  lower than 3 and consistently, for the H\,{\sc ii} regions we excluded the objects showing this ratio larger than 3. This is to avoid contamination by compact H\,{\sc ii} regions in the PN sample and to avoid contamination by supernova remnants, W-R nebulae and other highly excited objects in the H\,{\sc ii} region sample. Our diagrams then contain only genuine PNe and H\,{\sc ii} regions.

\medskip

The galactocentric distances employed to build the gradient diagrams  were calculated by us, according to the procedure described by \citet{cioni:09}.

\subsection{NGC\,300}

 The almost  face-on spiral NGC\,300 is the less massive galaxy of the all sample studied here and the only one out of the Local Group. Also it is the latest in terms of the Hubble type, SA(s)d. It is similar to M\,33 in several aspects.
Analysis of the H\,{\sc ii} regions have been made by several authors. \citet{bresolin:09} were the first to determine the oxygen abundance gradient based on {\it direct-method} abundance determinations, they found  $\Delta$O/$\Delta$R =  $-0.077$ dex kpc$^{-1}$ with a central abundance of 12+log O/H $\simeq$ 8.57. 

\citet{stasinska:13} analyzed the chemical gradients provided by H\,{\sc ii} region and PN abundances finding that gradients for PNe appear flatter than those of H\,{\sc ii} regions.

In this work we have re-analyzed the PN data presented by \citet{stasinska:13}, to make a comparison of the case of NGC\,300 with the other galaxies of the sample. Abundances presented by \citet{stasinska:13} were derived by adopting the ICFs proposed by \citet{kingsburgh:94}, and in this work  we have used the more recent ICFs presented by \citet{delgado-inglada:14}. With these new ICFs we found that our O/H values are equal to the \citet{stasinska:13} ones with differences lower than 0.03 dex, Ne/H values show  differences, in average, lower than 0.05 dex, while our Ar/H values are different from those by \citet{stasinska:13} by 0.07 dex in average. Data for H\,{\sc ii} regions are from \citet{bresolin:09}. 

In this galaxy PNe and  H\,{\sc ii} regions have been found up to $R/R_{25}\simeq 1$ (R$\sim$ 5 kpc). Galactocentric distances were collected from the same authors as in the abundances.

\subsection{Milky Way}

Metallicity gradients have been widely studied in the MW by means of H\,{\sc ii} regions, PNe, Cepheid stars, stellar clusters and other objects. See e.g.,  \citet{deharveng:00}; \citet{maciel:03}; \citet{henry:10};  \citet{stanghellini:18, stanghellini:10}; \citet{esteban:17};  the compilation by \citet{molla:19}, etc. The reported results for PNe have been contradictory. It has been claimed that the PN gradients  coincide with those of H\,{\sc ii} regions or  that the PN gradients are shallower, that gradients have  flattened (or steepened) with time, or that the gradient changes of slope at certain distance  from the galactic center. One of the main problem in these determinations is the large uncertainties in the distances to PNe.

Determining the distance to galactic PNe is a difficult task. The trigonometric parallax method is available only for  a handful of nearby objects. Even the parallaxes measured by GAIA are limited to distances of a few kpc around the solar system, and in GAIA Data Release 2 less than a hundred PNe have confident measured parallaxes \citep{kimeswenger:18}. Therefore, the distances for a large sample of PNe are based on model-dependent statistical methods which  in occasions lead to different results.  To the present the most used distance scales are those proposed by \citet{stanghellini:10}, hereafter S10,  and  by \citet{frew:16}, hereafter F16. The latter authors established a robust optical statistical distance indicator,  the H$\alpha$ surface brightness vs. radius (S$_{H\alpha}$-r) indicator, where the  intrinsic radius is calculated by using  the angular size, the integrated H$\alpha$ flux, and the reddening to the PNe. This radius, combined with the angular size, yields directly the distance.  On the other hand, S10 determine statistical distances based on apparent diameters and 5 Ghz fluxes of PNe. A comparison of both distance scales are presented further below.

 In this work we used PN chemical abundances reported by \citet{henry:04}, \citet{milingo:10}, and \citet{henry:10}. These authors belong to the same group therefore all physical conditions were calculated with the same systematical method and total chemical abundances were obtained using the ICFs described in \citet{kwitter:01}. 
The Galactic sample consists of 156 PNe covering galactocentric distances in the range 0.21 kpc $\leq$ R $\leq$ 22.73 kpc (0.02 $\leq$ R/R$_{25} \leq$ 1.97). More than 40 of these  PNe lie at distances larger than the solar galactocentric distance and are crucial for the gradient determination.

In the case of  H\,{\sc  ii} regions, oxygen abundances were taken from \citet{esteban:17}, while neon and  argon  abundances 
were collected from  \citet{garcia:04}, \citet{garcia:05}, \citet{garcia:06}, \citet{garcia:07},  \citet{esteban:04}, \citet{esteban:13}, and \citet{fernandez:17}. As said above, the use of data processed by different authors can introduce some  inhomonogeneities and uncertainties in the Ne and Ar gradients  of H\,{\sc  ii} regions, that should be considered  carefully.

In  Fig. \ref{fig:mw-mg} we present the comparison of the metallicity gradients of oxygen, argon and neon for the Galactic H\,{\sc  ii} regions and the PN sample. This figure will be discussed in detail in \S  3.4, here we want to show that for the case of PNe we are using  the distances by F16 and S10 for a comparison. As seen in this figure the oxygen, neon, and argon abundances present a very large dispersion at all galactocentric distances, but the linear regressions for PNe are similar for both distance indicators. Therefore in the following F16 distances will be used as an independent way to compare with other works.

\section{Abundance gradients}
%Section 3
\label{gradients}

Abundance gradients of the different elements for the different galaxies were calculated by fitting a straight line to the abundances versus the fractional galactocentric distance R/R$_{25}$ and also versus the distance R (kpc).
For each elemental abundance  we computed one fit for PNe and one for H\,{\sc ii} regions. Gradients are shown in the respective figures. In Table \ref{tab:lin-fit} a compendium of the  abundance gradients versus galactocentric distance  R (kpc) is listed. For each fit the table gives the values at the intercept, $X_0$, and the slope, calculated in the equation:

\smallskip
 $Y = X_0~+ ~\Delta X /  \Delta R ~\times$ R (kpc).

\noindent Errors have been calculated at 1 sigma. The errors in the gradients of Ne and Ar are much larger than in O, due to the large dispersion in the abundances at any galactocentric distance and due to the uncertainties in the abundance determination because large ICFs are used for these elements.

\subsection{M\,31}

\begin{figure*}
\centering
\includegraphics[scale=0.4]{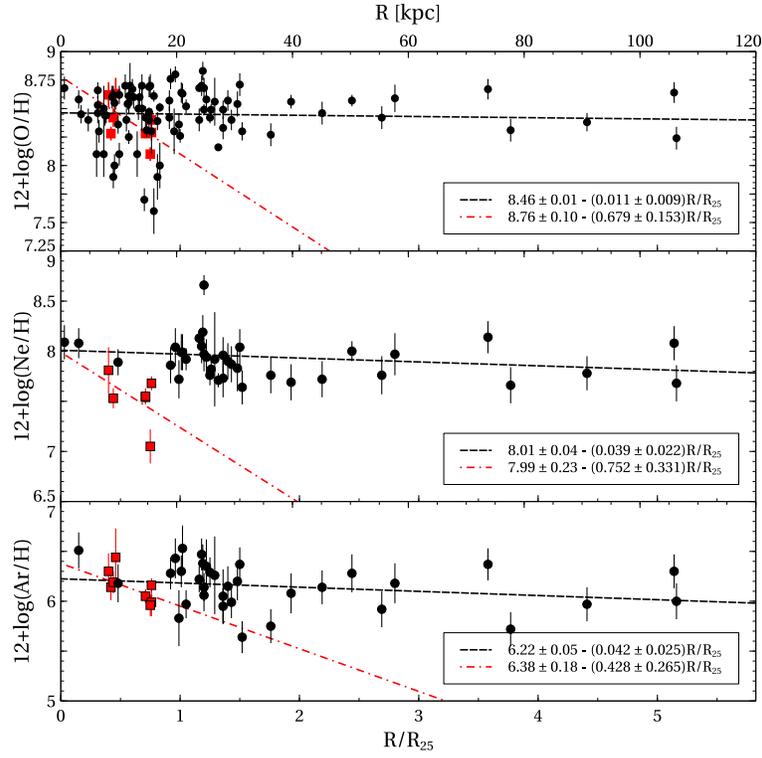}
\caption{Radial metallicity gradients of O (top), Ne  (center) and Ar (bottom) for M\,31, are presented. Black circles represent PNe and red squares, H\,{\sc ii} regions. Black dashed lines correspond to the linear fit for PN data and dashed-dotted red lines correspond to the linear fits for  H\,{\sc ii} regions respectively. \label{fig:M31-mg}}
\end{figure*}

 In Fig. \ref{fig:M31-mg} the radial gradients for O/H , Ne/H  and Ar/H , for PN and H\,{\sc ii} region abundances are presented vs. R/R$_{25}$. The gradients vs. R (kpc) are presented in Table \ref{tab:lin-fit}. 

A linear fit to the gradients is included in each case. The abundances of elements  in PNe present a large dispersion at any given galactocentric distance, but in particular in the central zone. It is worth to notice that there are some PNe in the central region with very low O/H abundance, which do not have Ne/H or Ar/H abundance determinations. H\,{\sc ii} regions also present a large dispersion in the elemental  abundances at any galactocentric distance \citep{zurita:12,sanders:12}.

\noindent For the case of O in H\,{\sc ii} regions, it is obtained: 

12+log(O)H) = (8.76$\pm0.10)  - (0.679\pm0.153)$ R/R$_{25}$   

\noindent or  equivalently

12+log(O/H) = (8.76$\pm0.10) - (0.030\pm$0.007)  R (kpc).

\noindent The O gradient found is equal within uncertainties to one derived by \citet{zurita:12}.

\noindent For Ne, we find 12+log(Ne/H) = (7.99$\pm0.23)  - (0.036\pm$0.016)  R (kpc),  
and for Ar, we find  12+log(Ar/H) = (6.38$\pm0.018) - (0.021\pm$0.013) R (kpc)  

\noindent In all the cases, the errors correspond to 1 sigma 

PN abundances seem not related to the galactocentric distance and the PN gradients are really flat, showing slopes of $-0.001\pm0.001$ dex kpc$^{-1}$ for O, $-0.002 \pm0.001$ dex kpc$^{-1}$ for Ne, and $-0.002\pm0.001$ dex kpc$^{-1}$ for Ar, which considering the errors, are  consistent with 0. 
This indicates that at large galactocentric distances PNe present in average the same O/H abundances that in the central zones and the same is true for Ar and Ne.

The O/H value at the intercept  for H\,{\sc ii} regions, 12+log O/H = 8.76$\pm$0.10, seems slightly larger  than the value for PNe, 12+log O/H = 8.46$\pm$0.03, while Ne/H and Ar/H central values are similar for H\,{\sc ii} regions and PNe, within uncertainties.
However, due to the negative gradients for H\,{\sc ii} regions, at large distances (R $\leq$ R$_{25}$,) O/H, Ne/H and Ar/H abundances  in PNe are always larger than abundances in H\,{\sc ii} regions.

The larger value of O at the intercept for H\,{\sc ii} regions seems an artifact due  to the limited sample by \citet{zurita:12} which moreover present large uncertainties. 
In addition, there are a large number of PNe with low O abundances in the central zone,  with no Ne and Ar measurement, that could be contributing to the low O/H central value for PNe. In this zone a very large dispersion is observed.
 
In the central zone the value at R=0 of log Ne/O = $-$0.77 in  H\,{\sc ii} regions  is about solar and discards the possibility indicated by \citet{zurita:12} of a large O depletion in dust grains. On the other hand, Ne in H\,{\sc ii} regions is similar to Ne in PNe in the central zone. Relative to Ar, log Ar/O\,(H\,{\sc ii}) = $-$2.38 and  log Ar/O\,(PNe) = $-$2.24 which correspond well with the solar or Orion values.

\begin{figure*}
\centering
\includegraphics[scale=0.4]{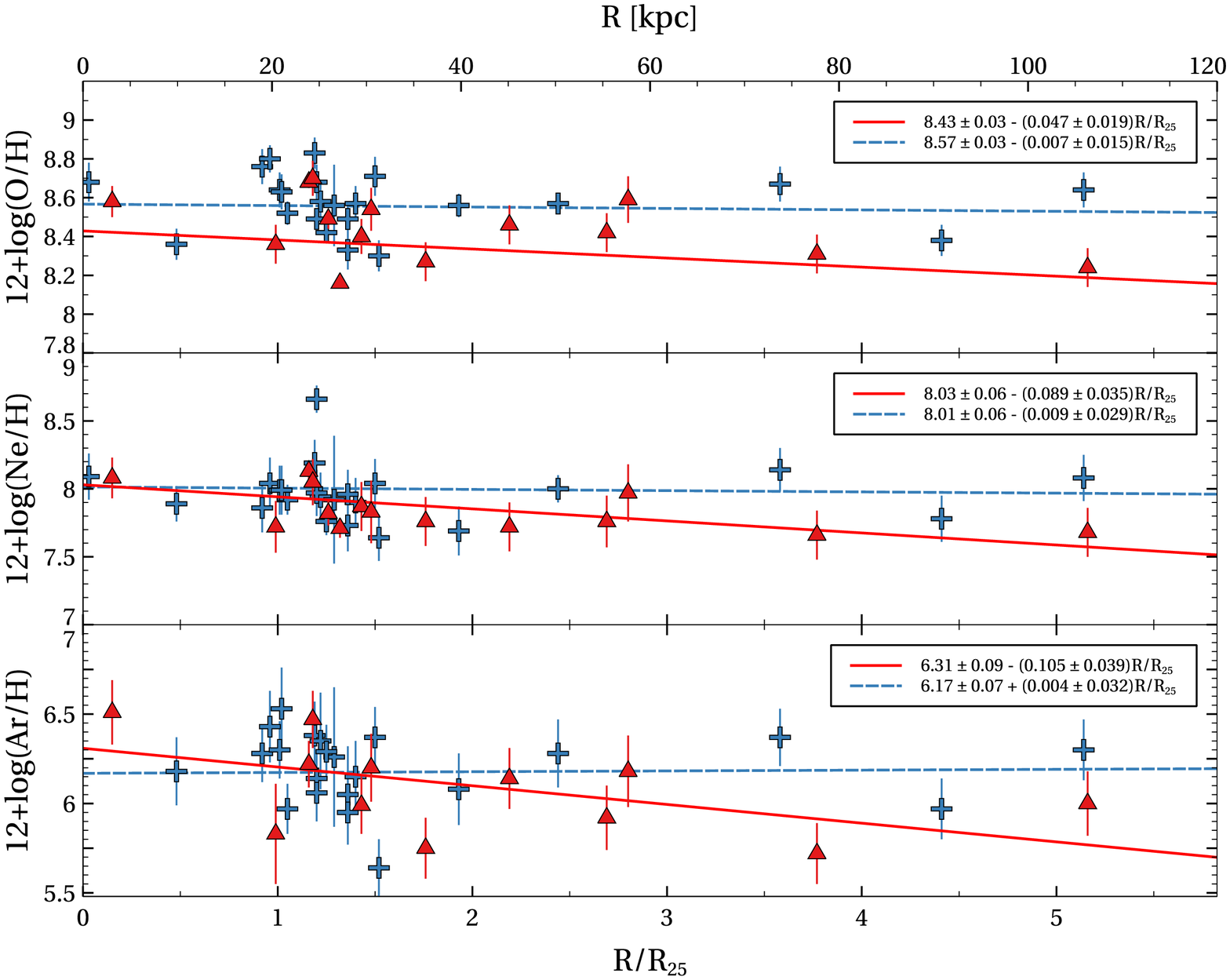}
\caption{Radial metallicity gradients for different Peimbert Type PNe in M\,31. In red the data for Type I PNe are shown and in blue, the data for non-Type I PNe. \label{fig:M31-PN-type}}
\end{figure*}

In Fig. \ref{fig:M31-PN-type}   we show the abundance gradients for PNe separated in Peimbert types. \citet{peimbert:78} called Type I  those PNe with  N/O abundance ratio larger than 0.5  and He/H abundance ratio larger than 0.14. This group  includes the PNe with central stars with initial mass larger  than about 3.0 M$_\odot$ and therefore they are the youngest objects among PNe. They have enriched their nitrogen abundances in nucleosynthesis  processes such as CNO and hot-bottom-burning (HBB) and the newly formed N has been transported to the surface in different  dredge-up events. In the Galaxy, Type I PNe belong to the thin disk and their ages are about 1 Gyr.  Non-Type I PNe include the Peimbert Types II and III PNe  which are classified based on their radial velocity, smaller or larger than 60 km s$^{-1}$, and are located in the thick disk. They correspond to older objects with initial masses lower than  2 M$_\odot$, where no such a  large N-enrichment is appreciated. Their ages are between 2 to 8 Gyr.

Thus our Fig.   \ref{fig:M31-PN-type} represents an effort to determine the behavior of the abundance gradient with time. It is evident that although the gradients are very  flat and the uncertainties are large, Type I PNe show  slightly steeper gradients for the three elements: O, Ne and Ar. Additionally, Type I PNe seem to be O-poorer than non-Type I, possible showing the effect of CNO and HBB processes  which operates in these massive stars.  This phenomenon will be discuss with more detail in \S 4.1. 

It should be mentioned here that the value for N/O abundance ratio, to define a Peimbert Type I  PN is slightly dependent on the metallicity. The N/O ratio defined by \citet{peimbert:78} applies to the metallicities of the Milky Way and M\,31, and should be slightly lower for M\,33 and NGC\,300, but the difference is not important for this work and the results are not much affected.

We will discuss the results for M\,31 in sections ahead, together with the results for the other galaxies. 

\subsection{M\,33} %S 3.2
\begin{figure*}
\centering
\includegraphics[scale=0.4]{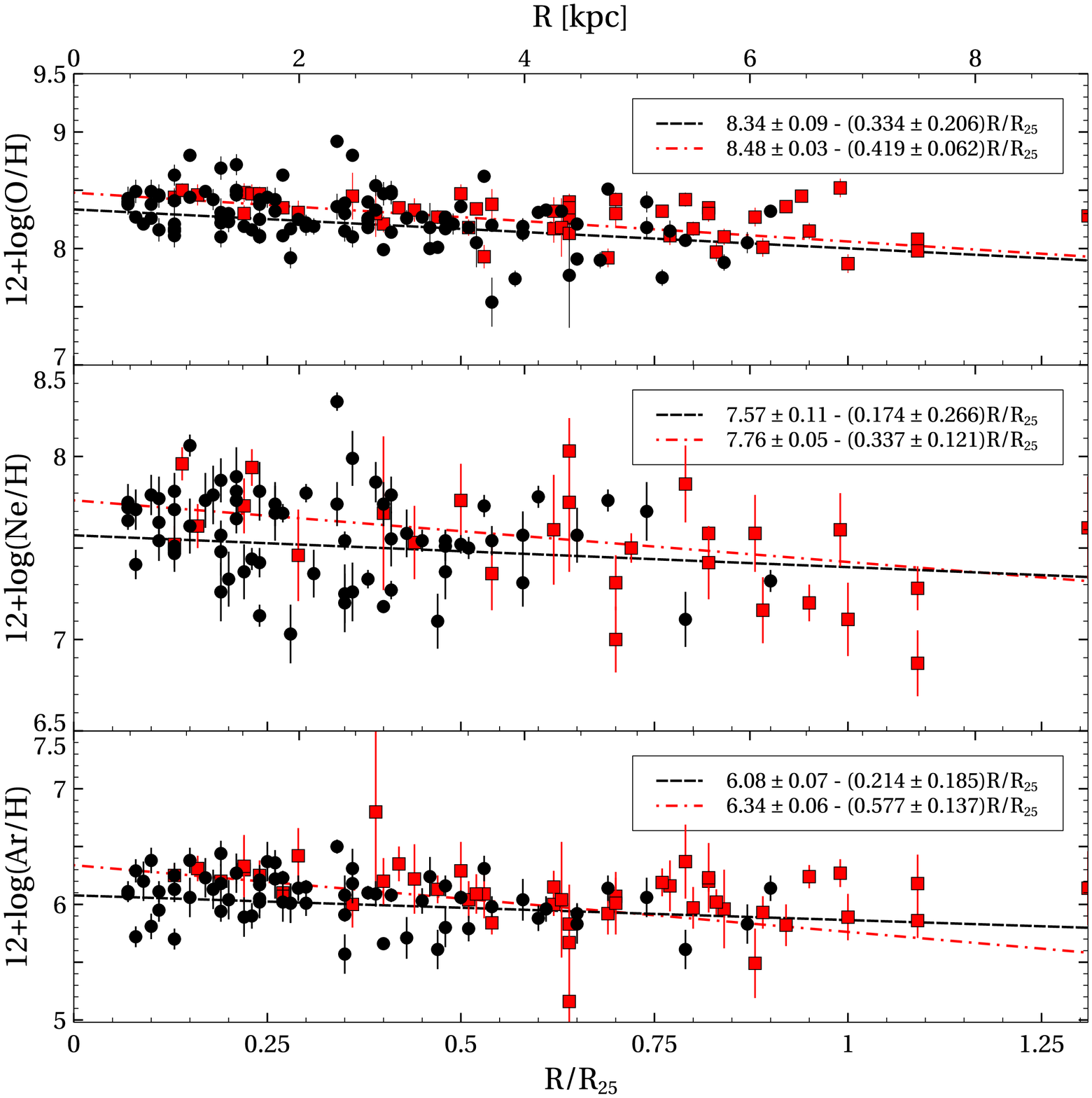}
\caption{ Radial metallicity gradients in M\,33. Symbols are as in figure \ref{fig:M31-mg}. \label{fig:M33-mg}}
\end{figure*}

\begin{figure*}
\centering
\includegraphics[scale=0.4]{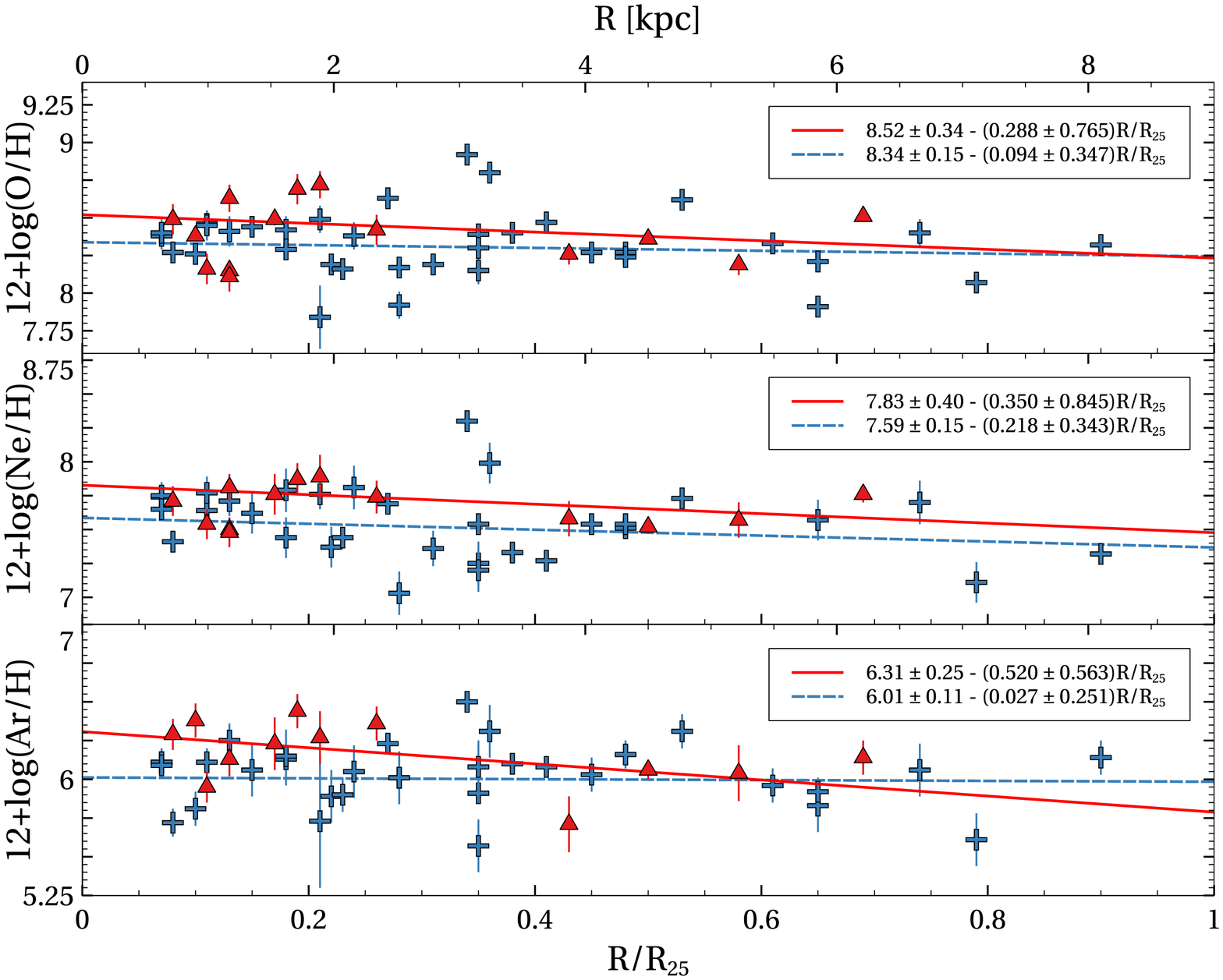}
\caption{Radial metallicity gradients from different Peimbert Types in M\,33. In red the data for Type I PNe are shown and in blue, the data for non-Type I PNe. \label{fig:M33-PN-type}}
\end{figure*}

Metallicity gradients for H\,{\sc ii} regions  and PNe in M\,33 are presented in Figs. \ref{fig:M33-mg} and \ref{fig:M33-PN-type}. In the latter one, PNe are separated in Peimbert Type I and non-Type I objects. 
Differently to what happens in M\,31,  in this galaxy O, Ne and Ar abundances of PNe and H\,{\sc ii} regions are similar within uncertainties, at any galactocentric distance.

For PNe  the straight-line fit of the gradients gives 12+log O/H = (8.34$\pm$0.07) $-(0.038\pm0.016$) R (kpc) and for  H\,{\sc ii} regions, 12+log O/H= (8.48$\pm0.03) - (0.047\pm0.008)$ R (kpc).  That is, the slope for PNe seems  shallower than the slope in H\,{\sc ii} regions, by about 30\% for O and larger factors for Ne and Ar,  but the uncertainties are large making these results not conclusive. However we consider indicative that PN gradients are shallower.

%At the intercept R=0, PNe present log Ne/O = $-$0.84 and log Ar/O = $-$2.27 which are similar to the solar values, and for H\,{\sc ii} regions we found  log Ne/O = $-$0.66, and log Ar/O = $-$2.15 which are again compatible with the solar values and also similar to Orion values.  This indicates that PNe have not  modified, through stellar nucleosynthesis,  their initial abundances  for O, Ne and Ar which are similar to values of the present ISM.

 In Fig. \ref{fig:M33-PN-type}, the gradients of PNe with different Peimbert Types are shown. There are only a few Type I PNe in the sample (about 20\%) and the uncertainties are very large, however the central values can be considered equal within errors, for both PN types. Despite the uncertainties, slightly steeper  gradients are apparent for Type I PNe, while  the gradients for non-Type I PNe are  flatter. The slopes shown by Type I PNe are more similar to the ones presented by H\,{\sc ii} regions.  The huge uncertainties in these results make them not conclusive, but only indicative. We consider them confident because M\,33 would be showing similar results to  M\,31 and NGC\,300.

\subsection{NGC\,300}
PN data in this galaxy were analyzed by \citet{stasinska:13} where O, Ne, S and Ar abundance gradients were presented. These authors found that the formal O/H, NeH and Ar/H abundance slopes  for PNe are shallower than those of H\,{\sc ii} regions and attributed this to a steepening of the metallicity gradients during the last Gyr. The O/H central value in PNe is smaller by 0.15 dex  than the central value of H\,{\sc ii} regions. Ne/H and Ar/H on the other hand present the same central abundances in PNe and H\,{\sc ii} regions, and almost flat gradients, although affected by large dispersion at any galactocentric distance. According to \citet{stasinska:13}, due to the difference in the O/H value at R=0,  O abundances in PNe could have been affected by nucleosynthesis of their central stars. 

\begin{figure*}
\centering
\includegraphics[scale=0.4]{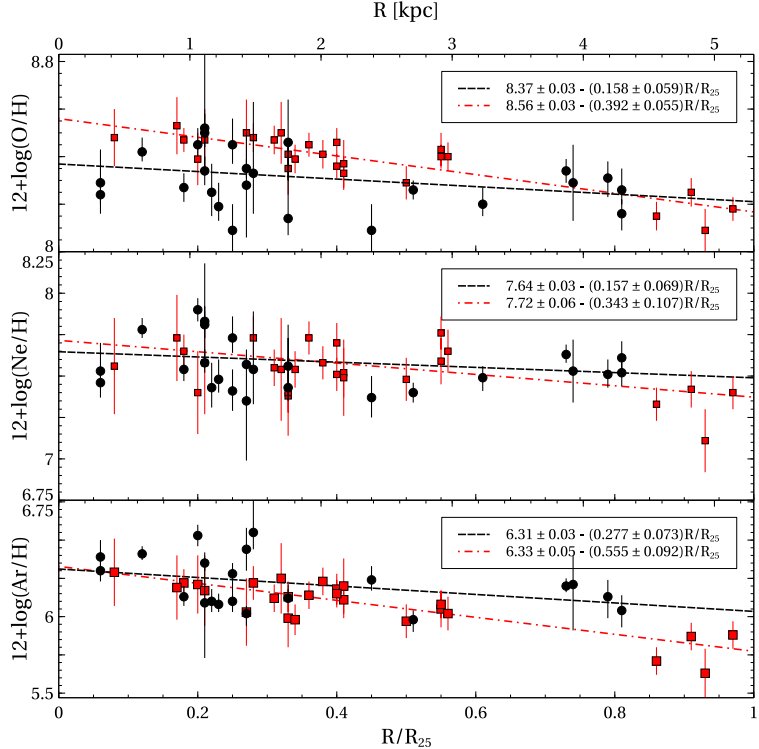}
\caption{Radial metallicity gradients in NGC 300. Symbols are as in figure \ref{fig:M31-mg}. \label{fig:ngc300-mg}}
\end{figure*}

Our analysis of these data, calculated with the ICFs by \citet{delgado-inglada:14} shows similar results  that are presented in Fig. \ref{fig:ngc300-mg}. O, Ne and Ar metallicity gradients for PNe in NGC\,300 are flatter than the values of H\,{\sc ii} regions. Their $\Delta  X / \Delta R $ values are about half the values found  for H\,{\sc II} regions, in very good agreement with the results by \citet{stasinska:13}. At the central zone,  H\,{\sc ii} regions show 12+log O/H = 8.57$\pm$0.03, log Ne/O = $-$0.86$\pm$0.008  and log Ar/O= $-$2.20$\pm$0.07, these abundance ratios are similar to solar or Orion values, while  PNe present values 12+log O/H= 8.37$\pm$0.03,   log Ne/O =$-$0.73$\pm$0.06, and log Ar/O = $-$2.06$\pm$0.05,  ratios also similar to the solar and Orion values but showing an apparent O decrease by 0.2 dex, relative to the O in H\,{\sc ii} regions, as  it was already indicated by \citet{stasinska:13}. This will be discussed in \S 4. The central values of Ne/H and Ar/H are similar for PNe and H\,{\sc ii} regions.

In this case, an analysis discriminating between the Peimbert Type PNe is not possible, because there are very few Type I PNe in the sample.

\subsection{The Milky Way}

As said above metallicity gradients have been widely studied in the MW by means of  many kind of objects. See references in \S 2.4.  

\citet{henry:10}, using distances by \citet{cahn:92} derived an O  gradient for PNe of $-0.058\pm0.006$ dex kpc$^{-1}$, which changes to $-$0.042$\pm$0.004 dex kpc$^{-1}$ if the distances by \citet{stanghellini:08} are used. \citet{henry:10} suggested that the gradient steepens beyond a galactocentric distance of 10 kpc. In a recent work, \citet{stanghellini:18}, using distances given by S10, reported that out to $R/R_{25}\simeq 2.4$  (R $\sim$  28 kpc) the radial gradient of oxygen for PNe is shallow, with slope $\sim -0.02$ dex kpc$^{-1}$ and a  central abundance of 12+log(O/H)$\simeq$ 8.68. 
These authors suggest that the gradient changes with R in the sense that the significant slope is limited to  R between 10 and 13.5 kpc and outside this range the gradient is almost flat.

 We analyzed the gradients derived from the data for PNe and H\,{\sc ii} regions mentioned in \S 2.4. PN  abundances were calculated by us in a homogeneous way. The results are presented in Figs. \ref{fig:mw-mg} and \ref{fig:MW.PNe.bin}. In the latter figure  the data have been binned in distance taking bins of 1 kpc, for clarity. The distances by F16 are used for PNe.

Clearly, PN gradients are flatter that those of H\,{\sc ii} regions. $\Delta$X/$\Delta$R is about twice larger for H\,{\sc ii} regions (see Fig. \ref{fig:MW.PNe.bin} and Table \ref{tab:lin-fit}). 
The binning of data with distances have introduced an artifact in Fig. \ref{fig:MW.PNe.bin} in the sense that O and Ne  abundances for PNe seem larger than the values for H\,{\sc ii} regions, at R=0. But this does not occur in Fig. \ref{fig:mw-mg} where the original data, not binned,  were used. In these graphs it is observed that the O and Ne 
 values at the central zones coincide for PNe and H\,{\sc ii} regions, within uncertainties, while Ar/H  abundance is lower in PNe by 0.7 dex. However due to the shallower gradients for PNe, at galactocentric distances larger than a few kpc, the average abundances in PNe are larger than in H\,{\sc ii} regions, similarly to what happens in M\,31 and NGC\,300.  By the way, this can explain part of the results by \citet{rodriguez:11} who found that in the solar vicinity, PNe appear richer than H\,{\sc ii} regions.

The gradients derived for PNe of different Peimbert types are shown in Fig. \ref{fig:MW.PNe.type}.  In this case, gradients of Type I and non-Type I PNe are equal within uncertainties. A possible change in the slope, at R$\sim$ 14 kpc, is appreciated that will be discussed in the next section.

We used the known GAIA  distances of PNe to analyze the gradient for PNe. 
Using the same set of abundances  of the MW, we searched for those PNe that have calculated parallaxes on GAIA Data Release 2 (GAIA DR2). We did not take into account those objects with negative parallax and error in parallax (dp/p) bigger than 0.4. We found 22 PNe meeting these requirements. No gradient is found in this interval for the three analyzed elements, because GAIA galactocentric distances are limited from 6 to 10 kpc, and this interval is too short to show any gradient, due to the large dispersion in abundances.

\section{Results and discussion}
\label{results} %Section 4
The radial gradients of the elements O, Ne and Ar are analyzed for homogeneous samples of PNe and H\,{\sc ii} regions in four disk galaxies of different Hubble type, different masses and different metallicities, M\,31, M\,33, NGC\,300 and the Milky Way. A compendium of our results, with all the linear fits and slopes $\Delta$X/$\Delta$R in dex kpc$^{-1}$, is presented in Table \ref{tab:lin-fit}. In the following we discuss the results.

\subsection{M\,31}

 Our work extends the PN sample, including objects from 2 kpc up  to a distance larger than 100 kpc (0.2 -- 5 R$_{25}$). In this interval the abundance gradients for PNe are flat, consistent with a slope of zero. 
The average abundances of PNe are the same at all galactocentric distances, showing a very large dispersion. 

On the other hand the gradients for H\,{\sc ii} regions are always negative, with values $-0.030\pm0.010$ dex kpc$^{-1}$ for O, $-0.036\pm0.016$  dex kpc$^{-1}$ for Ne and $-0.021\pm0.013$ dex kpc$^{-1}$ for Ar.  These slopes are much shallower than the slopes in the other three galaxies. The very flat gradients  found in M\,31 could indicate, according to \citet{sanchez:14}, that M\,31 has been perturbed by interactions or merging. It is clear that these phenomena have had an important role in the formation and growth of M\,31, a galaxy that shows numerous stellar substructures  in its outskirts \citep[and references therein]{mcconnachie:09}. 
 
At the central position PNe appear to have an average O/H abundance  slightly lower than the average in H\,{\sc ii} regions, but similar Ne/H and Ar/H abundances. This could be due to the presence of several PNe with very low O abundance (and not known Ne and Ar) in the central region, as it was explained in \S 3.1

The same as it is found in our work, \citet{sanders:12}  and \citet{magrini:16}  reported that the O/H gradient from PN abundances is flat in M\,31. \citet{sanders:12} in particular, from the analysis of O abundances in 52 PNe, derived with the {\it direct-method}, located at galactocentric distance from 5 to 25 kpc (0.2 to 1.2 R$_{25}$),  mentioned that there is no relation between PN abundances and their galactocentric distances.  This same PN sample was re-analyzed by \citet{magrini:16}  who reported that radial migration plays an important role in  PNe of M\,31, which possibly explains the extreme flatness of  PNe gradients, since  PNe may have migrated far from their  place of origin. 

When the sample is divided in Peimbert Type I PNe (young objects) and non-Type I PNe (older objects), it is found that the young objects show a steeper gradient, although still very flat (Fig. \ref{fig:M31-PN-type} ).  Due to their youth (ages lower than about 1 Gyr), Type I PNe have had less time to migrate from their  birth places, then they might be showing the gradient at about 1 Gyr ago, but considering that these gradients are very flat compared to the ones of H\,{\sc ii} regions,  
migration should have had an important role for these young PNe too.  

 It is important to mention that the results found for M\,31 corroborate models of galactic chemical evolution, which besides including the star formation rate, gas infall rate accross the disk, inflows and outflows of gas, stellar evolution and yields, among other processes, also include radial redistribution of stars (stellar migration). Models by \citet{ruiz:17}  predict that radial redistribution and accretion  increase the metallicity dispersion, and flatten the age and metallicity profiles of galaxies;  the greater the efficiency of the redistribution, the larger the flattening effect and, as a consequence, steeper metallicity gradient should be expected at birth of the objects.
Being Type I PNe closer to their birth places, they show steeper gradients.

Peimbert Type I PNe present slightly lower O abundances than non-Type I's, while Ne and Ar abundances are similar. This can be due to nucleosynthesis  because CNO and HBB processes  are expected to occur in these more massive central stars modifying their initial O abundance. Such an O decrease  is predicted by recent sophisticated evolutionary stellar models for  low-intermediate mass stars of the MW, computed  at different metallicities, by Ventura et al. (2017). Using the sequence of models with $z \sim 0.014$ (solar metallicity) it is  found that stars with masses larger than 3 M$_\odot$ diminish their O/H abundance by up to 0.2 dex at the end of their evolution, while stars of lower masses do not modify their original O/H. This is the effect we are finding in the comparison of Type I and non-Type I PNe in M\,31.

\subsection{M\,33}
In the case of M\,33, a late Hubble-type galaxy, the results are presented in Figs. \ref{fig:M33-mg} and \ref{fig:M33-PN-type}. In this case O, Ne and Ar in PNe show similar central values to the values of  H\,{\sc ii}  regions  therefore there is not important enrichment of the ISM since the time of formation of these PNe.  Similarly, the  O abundance gradients of PNe and H\,{\sc ii}  regions are equal, within uncertainties, but the Ne and Ar abundance  slopes in PNe seem  significantly flatter, although the large dispersion and  large uncertainties make these results doubtful.

Due to the similarity of metallicity gradients and the large dispersion, \citet{magrini:09} claimed that  gradients are equal for PNe and H\,{\sc ii} regions. In a recent paper, \citet{magrini:16} analyzed the possible effects of radial migration in M\,33 and concluded that it is not important. Also \citet{bresolin:10} declared that PNe and H\,{\sc ii} regions have equal gradients within uncertainties, but an analysis of their Table 8 shows that the slopes for PNe are systematically flatter than the slopes of H\,{\sc ii} regions, even when the uncertainties are large.

It is interesting to notice that PNe in M\,33 do not show O reduction compared to H\,{\sc ii} regions, as occurs in NGC\,300 (see next section), despite the similar low metallicity in both galaxies. It seems that the initial masses of the central stars in this galaxy are not as large as in NGC\,300, thus they are not affected by nucleosynthesis in the same way as in NGC\,300. According to \citet{ventura:16} models, the initial masses should have been not much larger than 2 M$_\odot$.

Although the uncertainties are huge due to the low number of objects, Type I PNe in M\,33 seem to present slopes steeper  than non-Type I PNe, and more similar to the ones of H\,{\sc ii} regions.  Again these results are very uncertain,  but we consider them indicative of the similar phenomenon ocurring in  M\,31.

\subsection{NGC\,300}

For NGC\,300, the latest Hubble-type galaxy, with metallicity similar to M\,33, the abundance gradients of PNe are about twice smaller than the gradients of H\,{\sc ii} regions (see Fig. \ref{fig:ngc300-mg}). 
The gradients presented by H\,{\sc ii} regions are the largest ($\Delta$O / $\Delta$R = -0.077$\pm$0.008 dex kpc$^{-1}$) in the whole galaxy sample.

The average O/H central abundance of PNe is lower than the value in H\,{\sc ii} regions by 0.2 dex, while Ne and Ar have the same central values. Such an O decrease, which does not occur in Ne and Ar, could be the result of stellar nucleosynthesis and dregde-up events. Stellar evolution models performed by \citet{ventura:16} for PNe in the SMC (which has a metallicity similar to the NGC\,300 one) predict a decreasing of the initial O occurring in stars with masses larger than 3 M$_\odot$, due to HBB, simultaneously a large N-enrichment occurs. Such N-enrichment is observed in the PNe of NGC\,300 \citep{stasinska:13}. Therefore we conclude that the central stars of PNe analyzed in this galaxy had in general large initial masses and are  younger than 1--2 Gyr. This is certainly due to a bias in the sample because only the brightest objects were observed at the NGC\,300 distance \citep{pena:12}.

The central values of Ne/H and Ar/H are similar for PNe and H\,{\sc ii} regions. Once again this indicates that PNe are young objects which had initial abundances similar to the present ISM.

An interesting fact is that at R/R$_{25}$ larger than 0.6, Ar/H abundances of PNe are definitely  larger than the values of H\,{\sc ii} regions, independent of the large dispersion. This is also found for Ne, but less marked. Since Ar it is not expected to have been modified by stellar nucleosynthesis of central stars, this is indicating that radial migration should have been important in PNe (despite their youth) and that PNe have changed their initial galactocentric distances, being churned in the galaxy, although not at the level of migration found in M\,31.

\subsection{The Milky Way}%section 4.4

\begin{figure*}%[!ht]
\centering
\includegraphics[scale=0.5]{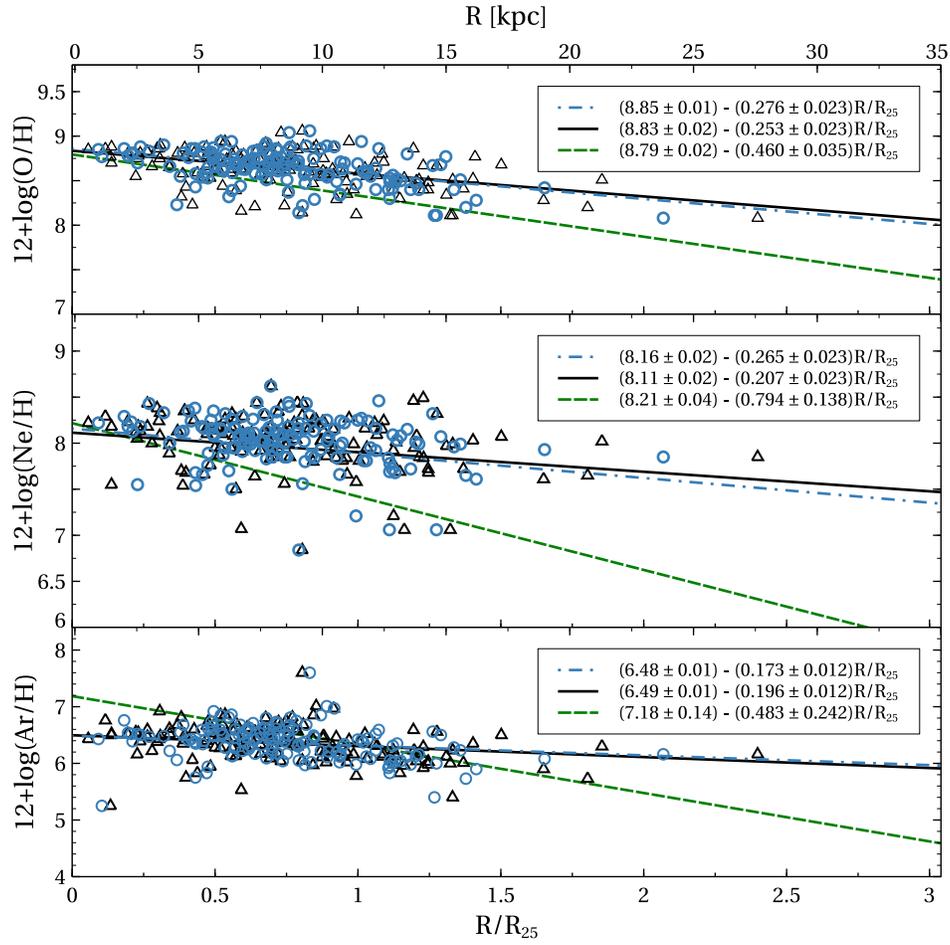}
\caption{Radial metallicity gradients of O, Ne and Ar, in the Milky Way, are presented. Open blue circles and black triangles are PNe with distances from  Frew et al. (2016) and  Stanghellini \& Haywood (2010) respectively. Dotted-dashed line is the linear fit for F16 distances and the solid line is for Stanghellini \& Haywood distances. Green dashed line is the  linear fit for H\,{\sc ii} regions by Esteban et al. (2016).  \label{fig:mw-mg}}
\end{figure*}

\begin{figure*}
	\includegraphics[width=\columnwidth]{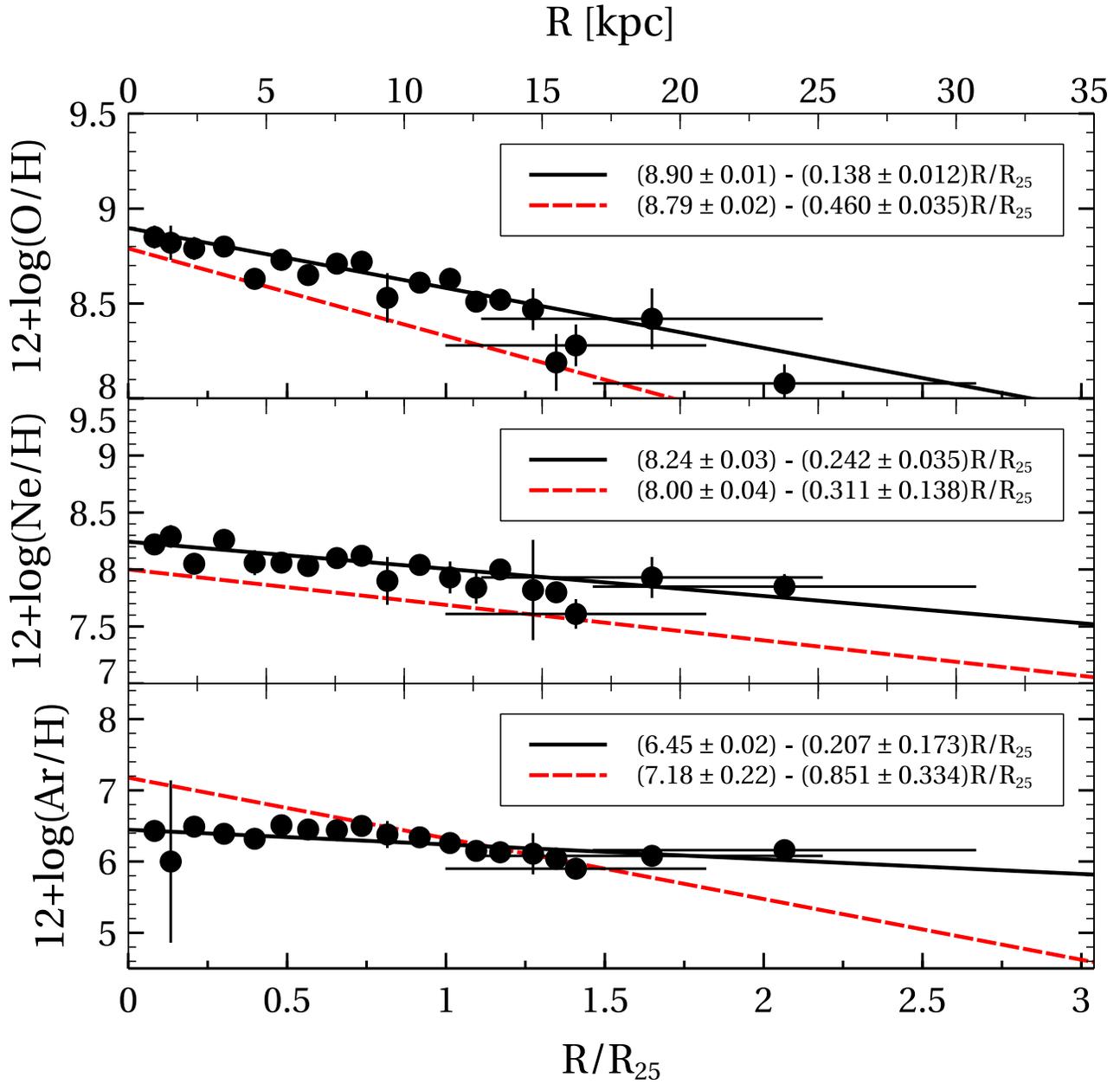}
    \caption{Black circles correspond to PNe with F16 distances binned in bins of 1 kpc. Continuous lines are the fits for PNe and dashed lines are the  fits for H\,{\sc ii} regions. On top we show the metallicity gradient for O, in the center the gradient for Ne, and at bottom the Ar metallicity gradient. }
    \label{fig:MW.PNe.bin}
\end{figure*}

The results for the MW are presented in Figs. \ref{fig:mw-mg}  and \ref{fig:MW.PNe.bin}. The latter figure shows PNe data with a bin of 1 kpc in distance.  The PN sample covers  a galactocentric distance interval 0.21 kpc $\leq$ R $\leq$ 22.73 kpc (0.02$\leq$ R/R$_{25} \leq$1.97). The gradient obtained for O in PNe is $-0.024$ dex kpc$^{-1}$, similar to the ones derived for Ne and Ar. As occurs in M\,31, and NGC\,300, PN gradients are flatter, by about a factor of 2,  than the gradients of H\,{\sc ii} regions.

\begin{figure*}
	\includegraphics[width=\columnwidth]{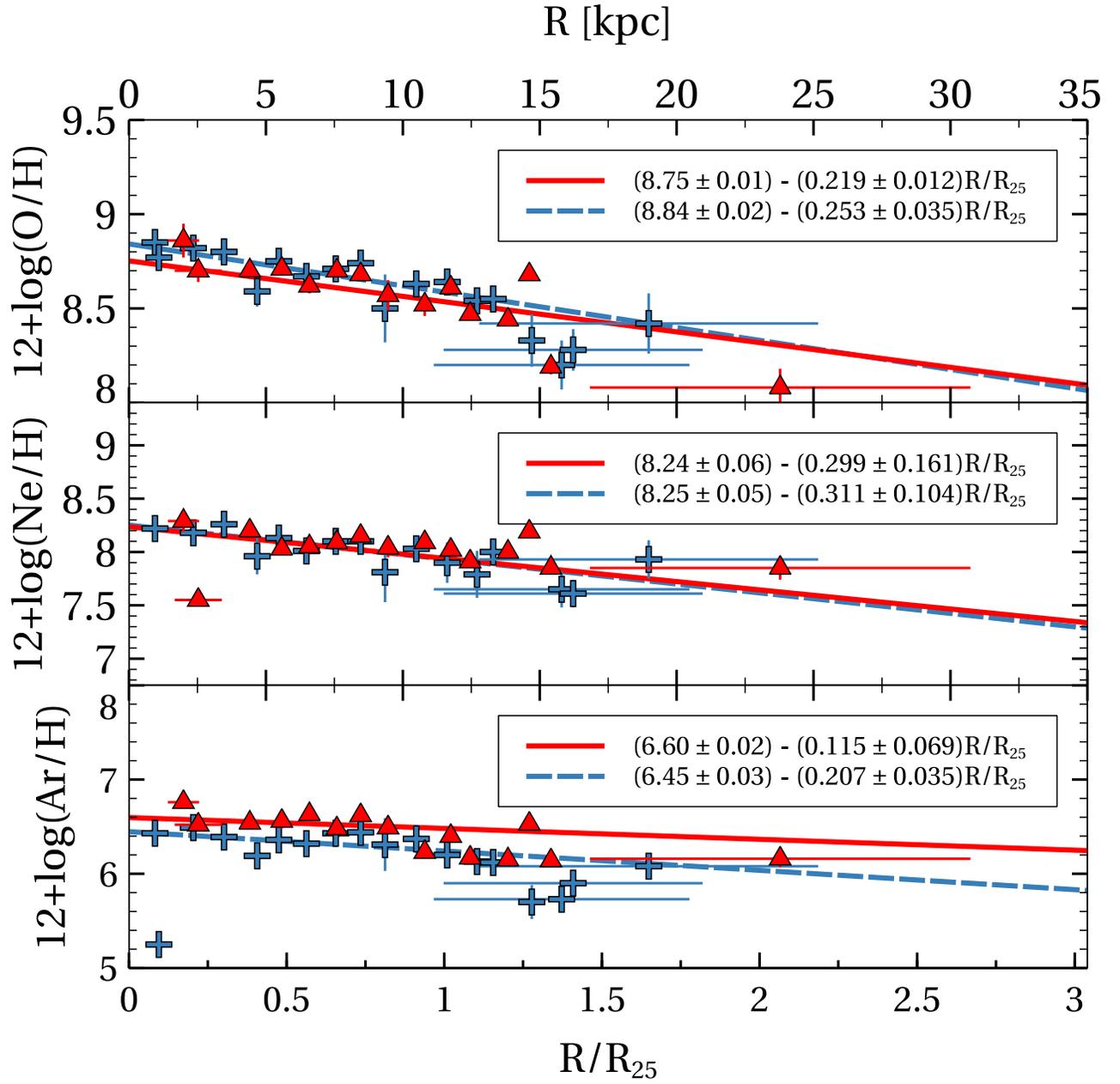}
    \caption{The PNe of the MW with same binned distances as in figure \ref{fig:MW.PNe.bin}. Red diamonds are Type I PNe and blue stars are non-Type I PNe. Continuous lines correspond to the linear fits for Type I's and the dotted lines correspond to fits for non-Type I's. }  
    \label{fig:MW.PNe.type}
\end{figure*}

The gradients for Type I and non-Type I PNe are shown  in Fig. \ref{fig:MW.PNe.type}.  In this figure the slopes of both kind of nebulae seem indistinguishable.    Our result is different from what  it was   reported by  \citet{stanghellini:10} and  \citet{stanghellini:18} for their young and old PN samples, using the distances by S10. They claim that young PNe definitely present a steeper gradient of $-0.027$ dex kpc$^{-1}$ versus $-0.015$ dex kpc$^{-1}$ for old PNe. We are not sure if our different result is due to the different PN samples (a much larger sample from the literature, not homogenized, was used by \citet{stanghellini:18})  or the different distance scale used. In any case we coincide with them  in that PN gradients are flatter than those of  H\,{\sc ii} regions, therefore   it appears than PNe have moved from their birth places due to radial migration and are showing flatter gradients. Alternatively it is possible that the abundance gradients were flatter several Gyr ago.

 Although a linear fit to these gradients produces acceptable results, it should be noticed that at distances larger than about 14 kpc, the observed abundances  (in particular O/H and Ar/H) decrease to a value below the linear fit and continue flat outwards, marking a step. Unfortunately in our sample there are  few PNe out of these distances and this result  is not conclusive. The same step at a distance R$\sim$13.5 kpc, was  reported by \citet{stanghellini:18} from the same sample of outer PNe. Following \citet{halle:15}  they attributed this behavior  to the effect of the galactic bar whose outer Lindblad resonance would be located at about this distance according to N-body  simulations for the galaxy. It is crucial to observe a larger number of PNe in the outskirts of the Galaxy  to verify this behavior.

It is interesting to compare the gradients of H\,{\sc ii} regions ($-$0.040 dex kpc$^{-1}$) and PNe ($-$0.024 dex kpc$^{-1}$) with others provided by well measured indicators of different ages. \citet{molla:19} have prepared a compilation of gradients for different objects of different ages, computed by different authors, in order to compare them with results from their chemical evolution models. From this compilation we select the gradients calculated by Moll\'a et al. for Cepheid stars which are young objects with ages of about  0.1 Gyr, showing an O/H gradient of $-$0.049 dex kpc$^{-1}$, and open clusters (OC) which span ages from 2 to more than 8 Gyr. The O/H gradient for OC with ages  younger than 2 Gyr is $-$0.030 dex kpc$^{-1}$, and for  ages between 2 and more than 8 Gyr, the slope is $-$0.027 dex kpc$^{-1}$. It is clear that the  gradient of H\,{\sc ii} regions and Cepheid stars are equal within uncertainties indicating that in the Galaxy  the chemical gradients have not changed significantly in at least the last 0.1 Gyr, while gradients of PNe (Type I's and non-Type I's)  are equal to that of OC with ages older than 2 Gyr.  Therefore for objects of 2--8 Gyr the gradient is flatter ($-$0.027 dex kpc${-1}$) than those of Cepheids and H\,{\sc ii} regions ($-$0.049 and $-$0.040 dex kpc$^{-1}$). PNe and OC certainly could have been affected by  migration, but not in an as extreme way as in M\,31. Alternatively, PN and OC gradients could represent the true gradients at several Gyr ago.  

In \citet{molla:19} it is discussed  the possible effects of migration in the MW. By analyzing models by different  authors, their conclusion is that radial migration seems to be not important for stars younger than 4 Gyr. Only for objects older than 8 Gyr, radial migration may be important.   Models by Moll\'a et al. of time evolution of chemical gradients in the Milky Way (without considering  migration, bar or spiral arms) predict a very smooth evolution of the radial gradient within the optical disk. Some model show a steepening of the gradient, from $-$0.02 to $-$0.04 dex kpc$^{-1}$ from time equal to 10 Gyr until the present. Therefore the gradients presented by PNe and OC could be the gradients at the time of formation of these objects, not affected by migration.

%figures for M\,33

%\citet{frew:16} \citet{stanghellini:10}

\smallskip

\begin{table*}
  \centering
  \caption{Linear fits for metallicity gradients in M\,31, the  Milky Way, M\,33, and  NGC\,300}
  \label{tab:lin-fit}
% {\scriptsize
%  \begin{tabular}{|r|c|c|c|c|}\hline 
\small
%\begin{changemargin}{-2cm}{-2cm}
	\begin{tabular}{lllc cc cc cc cc c}
\toprule
    & \multicolumn{4}{c}{Oxygen}  & \multicolumn{4}{c}{Neon}  & \multicolumn{4}{c}{Argon} \\
\cmidrule(l){2-5} \cmidrule(l){6-9} \cmidrule(l){10-13}
    &  $X_{0}$  & err & $\Delta X / \Delta R$    & err & $X_{0}$ & err & $\Delta X / \Delta R$    & err & $X_{0}$ & err & $\Delta X / \Delta R$    & err \\ 
  &                   &       & dex/kpc &   &    &   & dex/kpc &   &  &   & dex/kpc\\ 
\midrule

{\bf M 31}  &   & &  &  &  &  &  &   & & & &\\
All PNe  & 8.46 & 0.03 & -0.001 & 0.001 & 8.01 & 0.04 & -0.002 & 0.001 & 6.22 & 0.05 & -0.002 & 0.001 \\
 Type I  & 8.43 & 0.03  & -0.002 & 0.001 & 8.03& 0.05 & -0.004 & 0.002  &6.31 & 0.09& -0.005 & 0.002\\
non-Type I  & 8.57 & 0.03 &  -0.003& 0.001 & 8.01 & 0.06 & 0.000 & 0.001 & 6.17 & 0.07 & -0.000 & 0.001 \\
H\,{\sc ii} reg. & 8.76 & 0.10 & -0.030 & 0.007 & 7.99 & 0.23 & -0.036 & 0.016 & 6.38 & 0.18 & -0.021 & 0.013 \\
 \midrule
{\bf MW} &  &  &  &  &  &  &  &  &  &  &  \\
All PNe  & 8.85 & 0.05 & -0.024 & 0.003 & 8.20 & 0.03 & -0.021 & 0.005 & 6.58 & 0.03 & -0.018 & 0.015 \\
Type I & 8.75 & 0.05& -0.019 & 0.002 & 8.24 &0.06 & -0.026 & 0.014 & 6.60 & 0.02 & -0.010 &0.006 \\ 
non-Type I & 8.84 & 0.05 & -0.022 & 0.003 & 8.20 &0.05&  -0.027 & 0.009 & 6.45 & 0.03 & -0.018   & 0.006 \\
H\,{\sc ii} reg. & 8.79 & 0.05 & -0.040 & 0.005 & 8.21 & 0.04 & -0.027 & 0.012 & 7.18 & 0.14 &  -0.071 & 0.029 \\ 
\midrule
{\bf M 33} &   & &  &  &  &  &  &   & & & &\\
All PNe & 8.34 & 0.07 & -0.038 & 0.016 & 7.70 & 0.06 & -0.036& 0.030 & 6.17 & 0.06 & -0.031& 0.014 \\
Type I & 8.52 & 0.34 & -0.032 & 0.020 &7.83 & 0.40 & -0.039& 0.050 & 6.31 & 0.25 & -0.058 & 0.030 \\
non Type I & 8.34 & 0.15 & -0.010 & 0.020 & 7.59 & 0.15 & -0.024 & 0.015 & 6.01 & 0.11 & -0.010 & 0.020 \\
H\,{\sc ii} reg. & 8.48 & 0.03 & -0.047 & 0.008 & 7.76 & 0.04 & -0.043 & 0.010 & 6.34 & 0.04 & -0.064 & 0.016 \\ 
\midrule
{\bf NGC 300} &   & &  &  &  &  &  &   & & & &\\
PNe  & 8.37 & 0.03 & -0.030 & 0.011 & 7.65 & 0.03 & -0.029 & 0.013 & 6.31 & 0.02 & -0.051 & 0.014 \\
H\,{\sc ii} reg. & 8.57 & 0.03 & -0.077 & 0.008 & 7.71 & 0.05 & -0.065 & 0.016 & 6.33 & 0.04 & -0.104 & 0.017 \\
 \bottomrule
  \end{tabular}
%\end{changemargin}
\end{table*}
%  }

\section{Conclusions}

From the analysis of abundance gradients of O, Ne, and Ar in PNe and H\,{\sc ii} regions in the four galaxies, it is found that in NGC\,300 (a late Hubble-type spiral of low metallicity) and the Milky Way,  the abundance gradients for PNe are flatter than those of  H\,{\sc ii} regions, by factors of 2. This result is less conclusive in the case of M\,33, but also slightly  flatter gradients  are found for PNe than for H\,{\sc ii} regions. 

M\,31 represents an extreme case  where PN abundances are not related to the galactocentric distances and present the same values at any  distance from the center up to more than a 100 kpc. Merging, interactions with other galaxies and important radial migration of PNe are causing this phenomenon in M\,31. It is worth to notice that considering the H\,{\sc ii} regions,  M\,31 presents the shallower gradients of the four galaxies, which again is a possible consequence of merging and interactions.

In the four galaxies analyzed here, there is a large dispersion on abundances at any galactocentric distance which is  larger for PNe than for H\,{\sc ii} regions. This  possibly is caused by migration. PNe could have been churned in the galactic disk far from their birth places thus their abundances do not correspond to the place where they presently are. In trying to understand this it is important to analyze the abundances of elements that have been not modified by stellar nucleosynthesis, like Ar, as O and Ne can be modified depending on the metallicity and the stellar mass. However Ar  abundance determinations have large uncertainties, therefore the results based on Ar only should be taken carefully.

When considering the PNe separated by Peimbert types, M\,31 shows a clear case where the gradients seem to be steepening with time since  Type I PNe, which  correspond to the youngest objects among the PN population, with ages lower than 1 Gyr, present steeper slopes than non-Type I PNe, with ages between 2 to 9 Gyr. Type I PN gradients are more similar to, but still much flatter than H\,{\sc ii} region slopes. The gradients for non-Type I PN are most probably altered by radial migration that have moved the old PNe from their  initial galactocentric position in the galaxy. Type I PNe, due to their youth, have had less time to migrate from their place of birth, although the flatness of their gradients also indicate perturbations due to migration.  

In the Milky Way, H\,{\sc ii} region and PNe gradients can be compared to the gradients of objects of similar ages as Cepheid stars (age younger than 0.1 Gyr) and Open Clusters (ages between 2 and more than 8 Gyr). H\,{\sc ii} regions and Cepheid stars present the same O/H gradients, which indicates that the chemical enrichment in the Galaxy has not increased in the last few hundreds of Myr. On the other hand PNe and Open Clusters (OC)  present similar gradients, flatter than those of Cepheids and H\,{\sc ii} regions. This could indicate that the gradients steepen slowly with time (from $-$0.02 to $-$0.04 dex kpc$^{-1}$ in several Gyr). Alternatively it could indicate that radial migration could have perturbed the slopes presented by these relatively old disk objects. Chemical evolution models for the Galaxy by \citet{molla:19} seem to favor the first option.

In the Milky Way a sort of step in the PN gradient seems to exist in the sense that inside R $\sim$ 14 kpc, there is a measurable gradient of about $-$0.024 dex kpc$^{-1}$ in the three elements. Outside this galactocentric distance, PNe show a flat slope. \citet{stanghellini:18} reported the same phenomenon, based on the same data for PNe in the outskirts. However there are only a handful of objects detected  in this zone, and  a larger sample should be analyzed to confirm this step in abundances. 

In general, the central abundances of O, Ne and Ar of PNe are similar to the central abundances  of H\,{\sc ii} regions (the differences could be of about 0.2 dex), indicating that the central enrichment has not been important since the time of birth of PNe, and dust depletion of O in H\,{\sc ii} regions is not large. Although, due to the flat gradients of PNe, at galactocentric distance larger than about 0.5 R/R$_{25}$, the average abundances in PNe are larger than the average abundances in H\,{\sc ii} regions.

It is found that in PNe formed in low metallicity  environment ($Z \leq$ 0.004, like in NGC\,300), with initial masses of about 3 M$_\odot$ or larger, the oxygen could have been modified by stellar nucleosynthesis and its value in the nebula does not represent the initial value at the time of stellar formation. This is not the case in larger metallicity environments or in stars with lower initial masses. In the former case we suggest to use Ar as an element showing the initial abundance in the nebula, in spite of Ar abundance determination has large uncertainties.

 Our results are very useful for the computation of chemical  evolution models in these galaxies, as the present ISM represented by H\,{\sc ii} regions and the older component represented by PNe can be used for constraining the models. Non-Type I PNe are very useful to analyze the effect of radial migration in disk galaxies.

\section*{Acknowledgements}
This work received financial support from DGAPA-UNAM PAPIIT 103117 and  CONACyT Project 241732. S.N.F.-D. acknowledges postdoctoral scholarship from project CONACyT 241732. We acknowledge an anonymous referee for her/his careful revision and comments that help to improve this manuscript.

\end{document}